\documentclass[twocolumn,pra,showpacs]{revtex4}
\usepackage{graphicx}% Include figure files
\usepackage{amssymb}
\usepackage{amsmath}
\usepackage{float}
\usepackage{setspace}

\begin{document}

\title{Unified approach to topological quantum computation with anyons: From qubit encoding to Toffoli gate}

\author{Haitan Xu}
\author{J. M. Taylor}
\affiliation{Joint Quantum Institute and the National Institute of Standards and Technology, \\
College Park, Maryland 20742, USA}

\begin{abstract}
Topological quantum computation may provide a robust approach for encoding and manipulating information utilizing the topological properties of anyonic quasi-particle excitations.
We develop an efficient means to map between dense and sparse representations of quantum information (qubits) and a simple construction of multi-qubit gates, for all anyon models from Chern-Simons-Witten SU(2)$_k$ theory that support universal quantum computation by braiding ($k\geq 3,\ k \neq 4$).  In the process, we show how the constructions of topological quantum memory and gates for $k=2,4$ connect naturally to those for $k\geq 3,\ k \neq 4$, unifying these concepts in a simple framework. Furthermore, we illustrate potential extensions of these ideas to other anyon models outside of Chern-Simons-Witten field theory.
\end{abstract}

\maketitle

\section{introduction}
Current efforts in implementing ideas from quantum information
processing are largely focused on identifying systems with long
coherence times and the possibility of robust quantum control.  In principle, the topological properties of certain systems can be utilized to encode
and process information, which may protect the computation from local
noise~\cite{Kitaev03,Freedman02,Freedman03,Nayak07,Brennen08}.  In such systems,
low-energy excitations (quasi-particles) of highly correlated
two-dimensional systems can exhibit non-abelian anyonic statistics~\cite{Moore91,Read96,Kitaev06,Ardonne99,Ardonne01,Ardonne07,Cappelli99,Chung06,Fendley05,Levin05,Fu08,Qi09,Sau10,Nayak96}.
Braids of the non-abelian anyons in (2+1)-dimensional space-time correspond to fault-tolerant unitary
operations.

While the experimental implementation and observation of anyonic
systems remains an outstanding problem, anyons themselves present
additional challenges for building efficient computing devices, as they lack
the typical tensor-product structure associated with quantum bits.  In
the Chern-Simons-Witten SU(2)$_k$ theory, it was shown that anyon
models with $k =3$ or $\geqslant 5$ are universal for quantum
computation by braiding operations~\cite{Freedman02}. Explicit methods for
constructing qubits and single- or two-qubit gates for different $k$ have
been obtained in various ways in these anyon models except for the case $k=8$
~\cite{Bonesteel05,Hormozi07,Hormozi09,Xu08,Xu09,Xu092}.
Anyon models that do not support universal computation,
such as the $k=2$ case, may be made universal by adding non-braiding
operations~\cite{Freedman06,Bravyi05,Bravyi06,Bonderson09,Dassarma05,Georgiev06,Bonderson10}.
Investigating the case of $k=2$ also holds interest both in
applications for quantum
memory~\cite{Dassarma05,Freedman06,Georgiev06} and because of the
greater confidence in observing it in realistic physical
systems~\cite{Moore91,Read00,Ivanov01,Fu08,Qi09,Sau10,Willett07,Xu11}.

\begin{figure}
\begin{center}
\includegraphics[width=8.5cm]{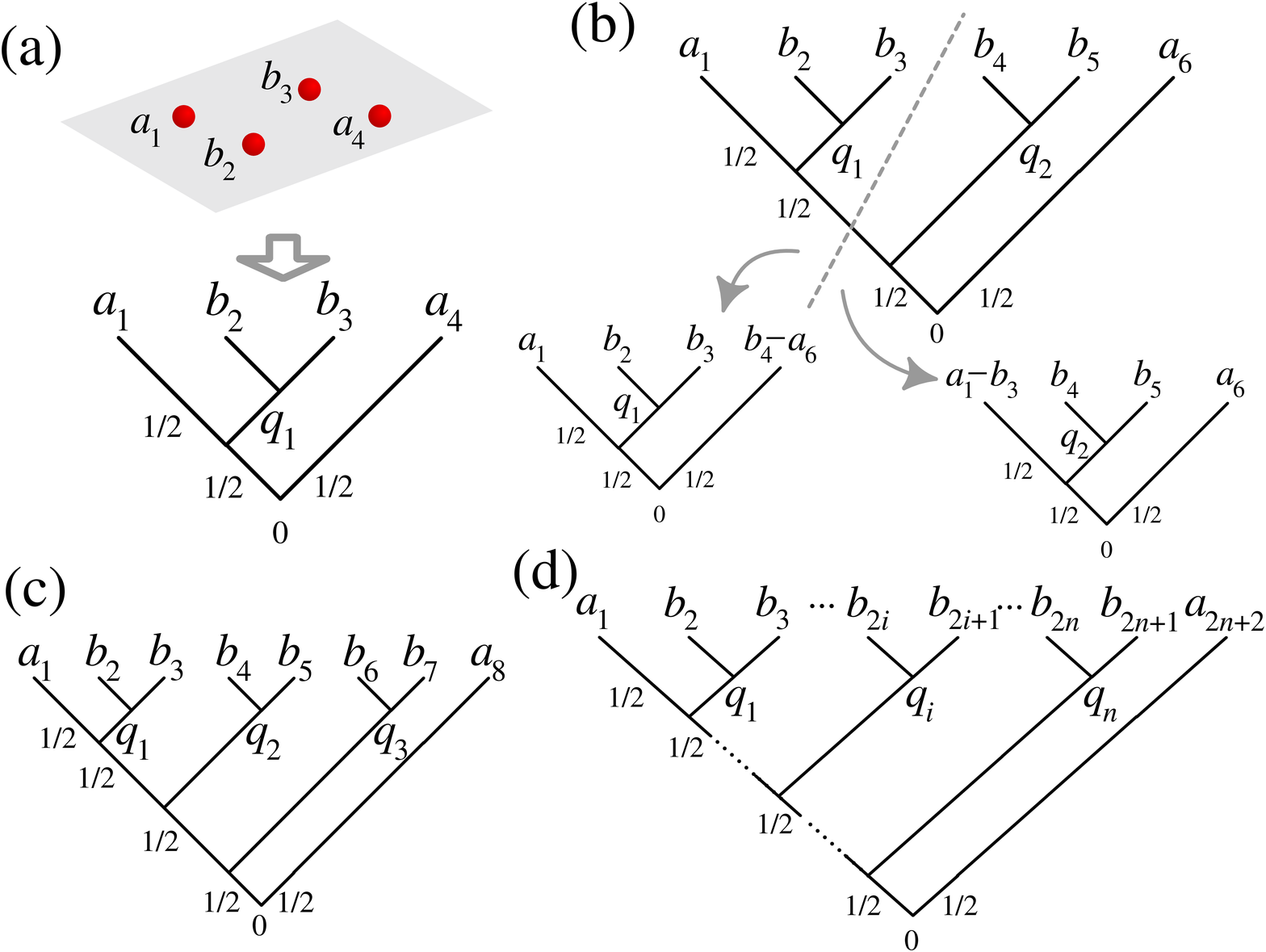}
\caption{\label{fig:groupfigure-01} (a) A single-qubit system composed of anyons in a two-dimensional structure, which we diagrammatically describe by physical anyons (labeled $a_i,b_i$) and lines indicating how their topological charges are combined.
(b) A two-qubit system composed of six anyons with total topological charge 0, where $q_1$ ($q_2$) is the total topological charge of $b_2$ and $b_3$ ($b_4$ and $b_5$), and specifies the first (second) qubit. By grouping together several anyons, we can map this system into two different single-qubit representations. (c) A three-qubit system composed of eight anyons with total topological charge 0. (d) An $n$-qubit system composed of $2n+2$ anyons with total topological charge 0.}
\end{center}
\end{figure}

In this paper, we develop a unified
framework for qubit encoding and multi-qubit gate construction in generic SU(2)$_k$ ($k\geq2$)
anyon models. We show the equivalence between different qubit
encoding schemes of prior works (e.g., \cite{Bonesteel05, Xu08, Georgiev06, Nayak96}).
For universal topological quantum computation (including $k=8$; note that we classify ``universal" or
``non-universal" topological quantum computation with regard to braiding operations), we give the construction of single-qubit and controlled gates.
We further show that the conservation of total topological charge and appropriate fusion rules allow us to aggregate information on different qubits in the total topological charge of composite anyons. This dramatically simplifies the construction of more complicated logical braiding operations. As an explicit example, we give braid topologies of controlled-controlled gates for universal topological quantum computation, which can improve the efficiency of actual computational operation in, e.g., Shor's algorithm.

\section{topological qubits and gates}
In a generic SU(2)$_k$ anyon model~(see, e.g., \cite{Fuchs92}), there are $k+1$ types of anyons, with topological charges $0, \frac{1}{2},...,\frac{k}{2}$. The fusion rule is $\frac{m}{2}\otimes \frac{n}{2} = \frac{|m-n|}{2}\oplus (\frac{|m-n|}{2}+1) \oplus \cdots \oplus \text{min} (\frac{(m+n)}{2},\frac{2k-(m+n)}{2})$. We can use four anyons ($a_1,b_2,b_3,a_4$) with total topological charge 0 to encode one qubit of information as in Fig.~\ref{fig:groupfigure-01}(a). The choice of topological charges for the anyons is not unique.  We choose $b$ such that the fusion rule leads to a non-trivial space, and $a$ such that the basis states from $b$ can add to total topological charge 0. Generally we can choose $a_i=b_j=\frac{1}{2}$.  To conform to prior work~\cite{Bonesteel05,Hormozi07,Hormozi09,Xu08,Xu09,Xu092}, in the SU(2)$_3$ model, we can also instead use Fibonacci anyons ($a_i=b_j=1$). For the SU(2)$_8$ model, we choose $a_i=\frac{1}{2}$, $b_j=1$ for convenience. 
We write the basis states of the single-qubit space as
$|0\rangle=| ( (a_{1} (b_{2} b_{3})^0)^\frac{1}{2} a_{4} )^0 \rangle$
and
$|1\rangle=| ( (a_{1} (b_{2} b_{3})^1)^\frac{1}{2}  a_{4} )^0 \rangle$, where the superscripts specify the total topological charges of the preceding brackets.
We can also equivalently use three anyons ($a_{1}, b_{2}, b_{3}$) to encode a single qubit, with basis states
$|0\rangle=| (a_{1} (b_{2} b_{3})^0)^\frac{1}{2}\rangle$
and
$|1\rangle=| (a_{1} (b_{2} b_{3})^1)^\frac{1}{2} \rangle$.

To encode two qubits, we can use six anyons with total charge 0 as shown in Fig.~\ref{fig:groupfigure-01}(b). The whole space of the six-anyon system is 5-dimensional (for $k>2$) or 4-dimensional (for $k=2$) , with computational basis states chosen to be
\begin{equation}
\label{eq:computationalbasis}
| q_1 q_2  \rangle = |   ( ( (a_1 (b_2 b_3)^{q_1})^\frac{1}{2} (b_4 b_5 )^{q_2})^\frac{1}{2}   a_6)^0     \rangle,
\end{equation}
where $q_{1,2}=0$ or $1$, and a non-computational basis state (for $k>2$)
\begin{equation}
\label{eq:non-computationalbasis}
| NC  \rangle = |   ( ( (a_1 (b_2 b_3)^{1})^\frac{3}{2} (b_4 b_5)^{1})^\frac{1}{2}   a_6)^0     \rangle.
\end{equation}
Here $q_1$ ($q_2$) is the total topological charge of $b_2$ and $b_3$ ($b_4$ and $b_5$), which specifies the first (second) qubit. Within the computational space spanned by computational basis, the total charge of anyons $b_4$-$a_6$ ($a_1$-$b_3$) is $\frac{1}{2}$, and can be treated as a single composite anyon, so that the two-qubit system is mapped to two single-qubit systems, as shown in Fig.~\ref{fig:groupfigure-01}(b). Though leakage into non-computational space can cause error, we will see the non-computational space plays an important role in the construction of controlled gates for universal topological quantum computation. Similarly, we can use eight anyons to encode three qubits as in Fig.~\ref{fig:groupfigure-01}(c).

Before going on to $n$ qubits, let us construct topological quantum gates in the above small systems. For {\it non-universal} topological quantum computation, we can search for gates directly by brute force as in~\cite{Georgiev06}, due to the finite-group nature of the allowed gates by braiding.   For {\it universal} topological quantum computation, things become more complicated.  Single-qubit gates, which are SU(2) matrices in single-qubit space, can still be obtained to arbitrary accuracy by search~\cite{Bonesteel05,Kitaev97Dawson06,Xu08,Xu09,Burrello10}. For example, the error rate of a generic single-qubit gate can practically be reduced to $\sim10^{-6}$ (or $10^{-10}$) with braid length $\sim100$ (or $300$) for Fibonacci anyon model~\cite{Xu08,Xu09,Xu092}.
For universal topological quantum computation, direct searching becomes prohibitive for the corresponding braids of multi-qubit gates due to the large parameter space of potential operations. Instead, we map a multi-qubit gate to series of operations acting effectively on two-dimensional Hilbert spaces, combined with appropriate transformations that restrict the multi-qubit space to appropriate subspaces.

\begin{figure}
\begin{center}
\includegraphics[width=8.5cm]{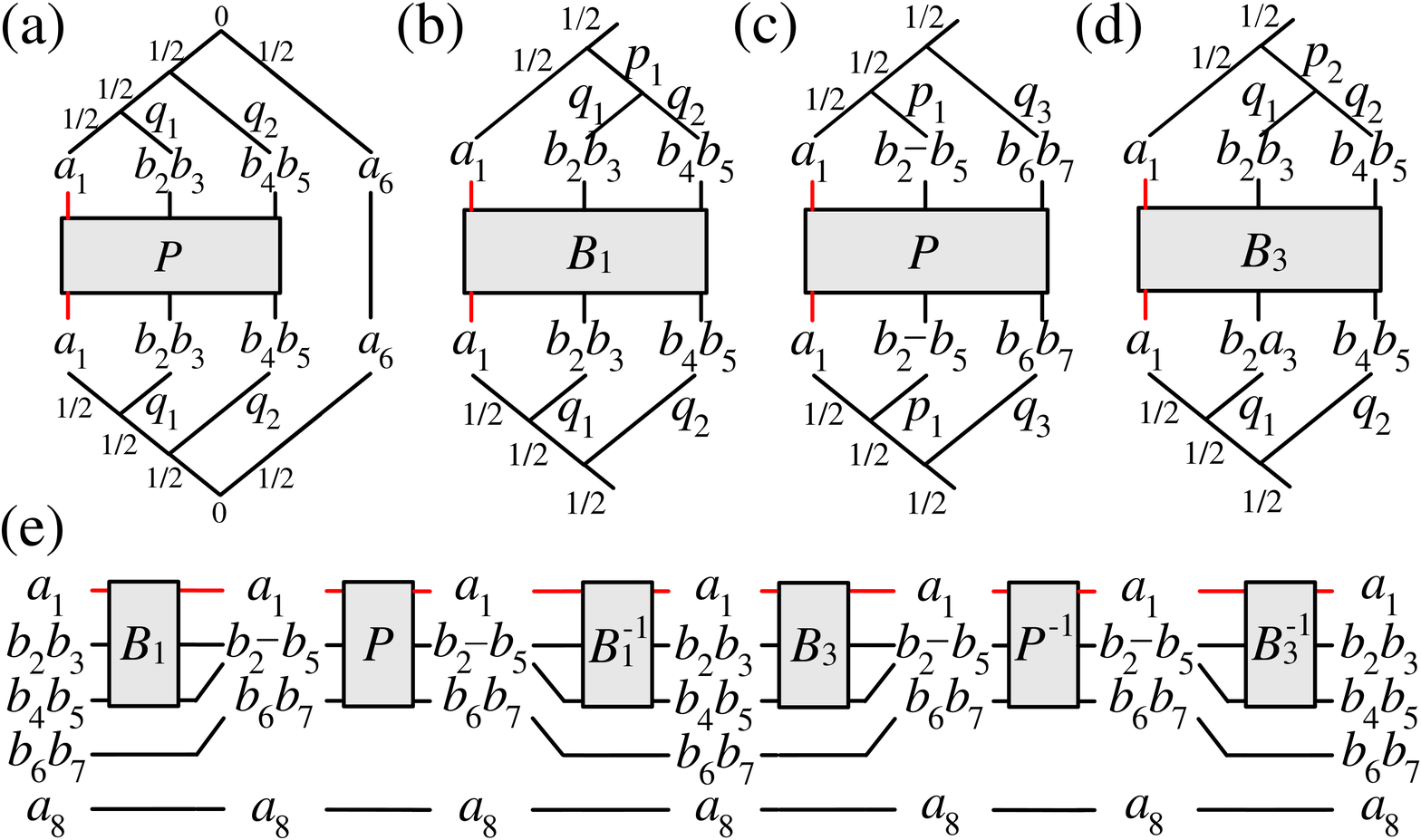}
\caption{\label{fig:groupfigure-03} (a) Controlled-phase gate, which is equivalent to a single-qubit phase gate $P$. See text for details.  (b) Combination of the information of $q_1$ and $q_2$ by a braid $B_1$, where we braid, e.g., $a_1$ around $q_1$ and $q_2$. We require the total charge of $q_1$ and $q_2$ to be 1 after performing the braid except for some arbitrary phase factor when $q_1=1$ and $q_2=1$. (c) A braid the same as the controlled phase gate in (a), except for different composite anyons. (d) A braid that changes the total charge of $q_1$ and $q_2$ to 0 when $q_1=1$ and $q_2=1$, except for some arbitrary phase factor. (e) Controlled-controlled-phase gate as described in the text. }
\end{center}
\end{figure}

We first show how to construct {\it controlled-phase gates} for universal topological quantum computation. Let us treat the anyons $b_2$ and $b_3$ as a composite anyon $q_1$, and also $b_4$ and $b_5$ as a composite anyon $q_2$, as shown in Fig.~\ref{fig:groupfigure-03}(a). This notion of composite anyons greatly reduces the space of braiding operations while preserving the explicit presentation of quantum information (represented by $q_1$ and $q_2$) during information processing, and thus simplifies the braid construction for the controlled gates. The whole space is separated into two sectors, i.e., the trivial cases where at least one of the composite anyons has topological charge 0, and the nontrivial case where both the composite anyons have topological charge 1. For the nontrivial case, the two-qubit system is in the state $|11\rangle$ (i.e., $q_{1,2}=1$), and $a_1$, $q_1$, $q_2$ and $a_6$ just form a two-dimensional Hilbert space, equivalently a single qubit, with basis states chosen to be
$| ( (a_1 q_1)^\frac{1}{2} q_2)^\frac{1}{2}  a_6 )^0 \rangle$ (within computational space)
and
$| ( (a_1 q_1)^\frac{3}{2} q_2)^\frac{1}{2}  a_6 )^0 \rangle$ (out of computational space). We can braid, e.g., $a_1$, around the composite anyons $q_1$ and $q_2$.  This braiding operation, involving three (composite) anyons, nominally appears to be a single-qubit gate.
Generally, such braiding operations will result in the so-called leakage error, which leads the computational states to non-computational states. Thus we require that the equivalent single-qubit gate be a phase gate, represented by a diagonal matrix in the basis $| ( (a_1 q_1)^\frac{1}{2} q_2)^\frac{1}{2}  a_6 )^0 \rangle$
and
$| ( (a_1 q_1)^\frac{3}{2} q_2)^\frac{1}{2}  a_6 )^0 \rangle$. 
Returning to the trivial cases, if the original two-qubit system is in the state $|00\rangle$ (i.e., $q_1=0$, and $q_2=0$), then braiding $a_1$ around charge zero composite objects ($q_1$ and $q_2$) will result only in a trivial phase 1, and the two-qubit system stays in the same $|00\rangle$ state. If the original two-qubit system is in the state $|10\rangle$ (or $|01\rangle$), there will be a phase determined by the total number of exchanges between $a_1$ and $q_1$ (or $q_2$). We require these phases to be 1, which imposes a constant difficulty in searching for the desired braid $P$ since all possible phases form only a small finite set.

Thus to construct a controlled-phase gate, we can search for a single-qubit phase gate $P$ under the weak restriction that the total number of exchanges between $a_1$ and $q_1$ (and between $a_1$ and $q_2$) is congruent to 0 modulo some integer (which depends on the anyon model), which can be easily found by computer-aided searching~\cite{Xu08,Xu09,Xu092,Burrello10,Kitaev97Dawson06}.  When the nontrivial (computational) matrix element of $P$ is -1, the corresponding controlled-phase gate is the controlled-Z gate.

This construction can be generalized to a variant of the Toffoli gate, the {\it controlled-controlled-phase gate}, which is of particular importance as an efficient method for performing many quantum algorithms such as the quantum Fourier transform. The main idea is to reversibly store the result of binary functions of two qubits into a single composite anyon, and then use this composite anyon to control the third qubit. Specifically, we work with a three-qubit system composed of eight anyons with total charge 0 (see Fig.~\ref{fig:groupfigure-01}(c)).
Let us treat the anyons $b_2$ and $b_3$ as a composite anyon $q_1$, $b_4$ and $b_5$ as a composite anyon $q_2$, and also $b_6$ and $b_7$ as a composite anyon $q_3$.  Braids will be composed of operations in which the composite anyons are braided collectively.  Our first goal is to find a braid $B_1$ that allows us to aggregate information about $q_1$ and $q_2$ in their total topological charge.  This can be accomplished by braiding $a_1$ around $q_1$ and $q_2$ (Fig.~\ref{fig:groupfigure-03}(b)).  The braid is chosen such that for $q_1$ or $q_2$ non-zero, the total charge of the composite anyon made by joining $q_1$ and $q_2$ is $p_1=1$, while for the case when both are zero, the total charge $p_1=0$, except for some phases that will be canceled later.  By separating the braid operation into the different topological charge subspaces associated with $q_1$ and $q_2$, we are confronted with one non-trivial braid ($q_1=q_2=1$ leading to $p_1=1$), which requires a search in a two-dimensional Hilbert space. 

Next we carry out a controlled-phase gate with the composite anyon $p_1$, which contains the information on the first two qubits, and $q_3$, which contains the information on the third qubit. We can directly use the gate $P$ in the earlier discussion of controlled-phase gate by mapping $p_1,q_3$ in Fig.~\ref{fig:groupfigure-03}(c) to $q_1,q_2$ in Fig.~\ref{fig:groupfigure-03}(a). 
After this, we carry out an inverse braid of $B_1$, which returns the total charge of $a_1$ and $q_1$ back to $\frac{1}{2}$, and cancels the phases 
introduced by the braid $B_1$.

We continue to integrate the information on the first two qubits in a different way. We perform a braid $B_3$, braiding $a_1$ around $q_1$ and $q_2$, as shown in Fig.~\ref{fig:groupfigure-03}(d), which changes the total charge of $q_1$ and $q_2$ to $p_2=0$ when $q_1=1$ and $q_2=1$, except for some arbitrary phase factor. When $q_1=0$ and $q_2=1$ (or $q_1=1$ and $q_2=0$), $p_2$ must be 1 according to the fusion rule. When $q_1=0$ and $q_2=0$, $p_2$ must be 0. Then we apply the inverse of the gate $P$ to cancel the controlled phase for the cases when $q_1$ and $q_2$ are not both 1. At last, we perform an inverse braid of $B_3$, which returns the total charge of $a_1$ and $q_1$ back to $\frac{1}{2}$, and also cancels the phases introduced by the braid $B_3$.

After performing all the six braids, there is a phase factor, e.g., -1, if $q_1$ and $q_2$ are both 1, and otherwise only a trivial phase 1. Thus we have obtained a controlled-controlled-phase gate with any phase, e.g., the controlled-controlled-
Z gate.
The whole braid sequence is summarized in Fig.~\ref{fig:groupfigure-03}(e). Note that we need to move only the anyon $a_1$ throughout the whole braid. $P^{\pm1}$ are equivalently single-qubit gates with three degrees of freedom fixed, and $B_{1,3}^{\pm1}$ are equivalently single-qubit gates with only two degrees of freedom fixed. The error rate of the gates can thus be reduced to the order of $10^{-10}$, with the number of elementary braiding operations between $a_1$ and composite anyons being of the order 700 or less (dependent on $k$).

\begin{figure}
\begin{center}
\includegraphics[width=8.0cm]{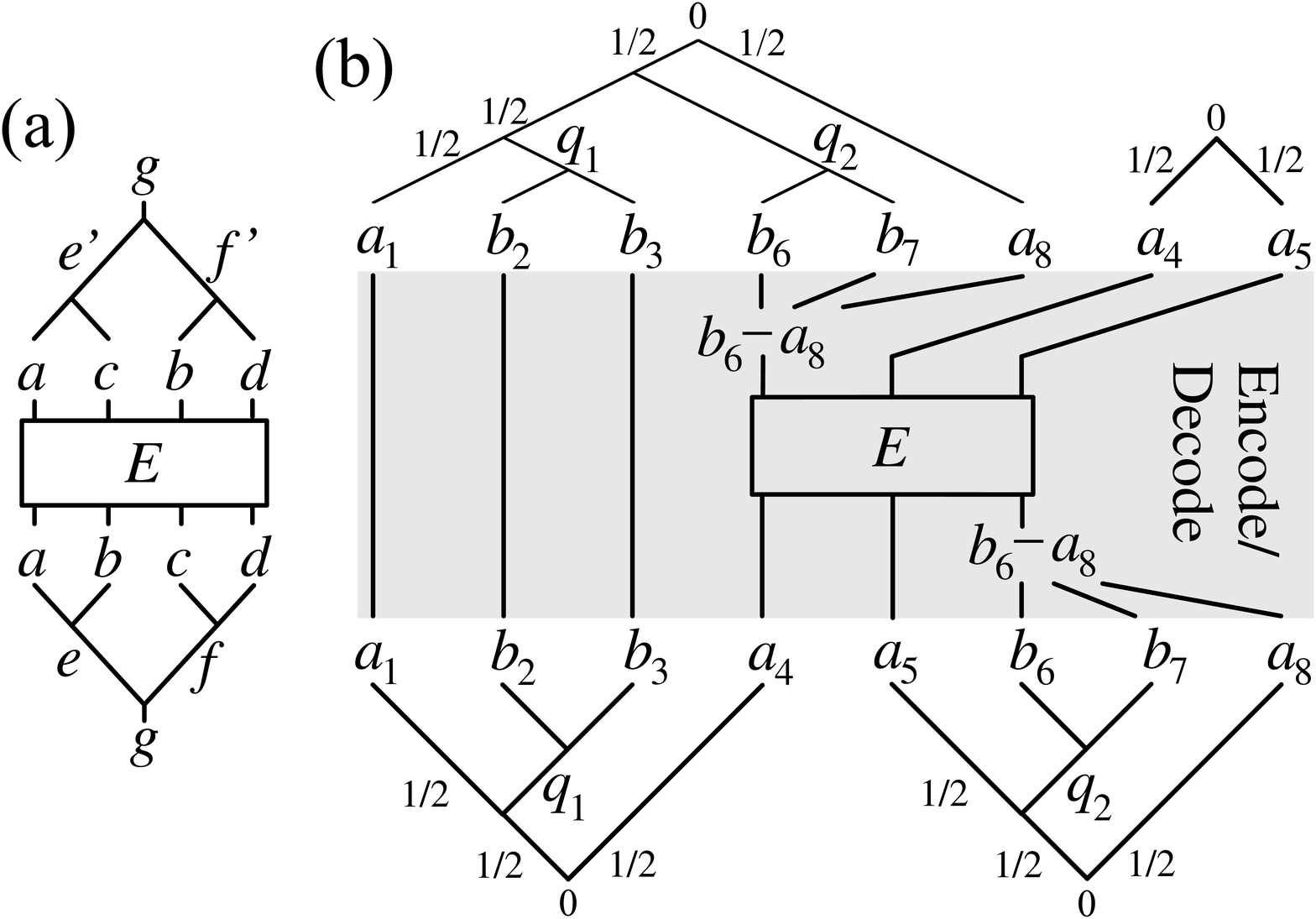}
\caption{\label{fig:groupfigure-02} (a) Anyon-exchange gate, which is a generalization of the exchange braid in Refs.~\cite{Xu08,Xu092}. $a$,$b$,$c$, and $d$ represents (composite) anyons that can have different topological charges, with total topological charge $g$. Originally the total topological charge of $a$ and $b$ (or $c$ and $d$) is $e$ (or $f$). The anyon-exchange gate exchanges the anyon $b$ and $c$, so that after the anyon-exchange gate, the total topological charge of $a$ and $c$ (or $b$ and $d$) becomes $e'$ (or $f'$). Such a gate can generally be realized by braiding or by projective measurement~\cite{Xu092}. (b) Encoding and decoding of the information in two single-qubit (four-anyon) systems into a two-qubit (six-anyon) system using anyon-exchange gate $E$. In this case, the anyon-exchange gate exchanges the anyon $a_4$ and composite anyon $b_6$-$a_8$ so that after performing the exchange gate, the total topological charge of $a_{1,8}$ and $b_{2,3,6,7}$ becomes 0, vice versa.
}
\end{center}
\end{figure}

A natural question that arises is how to embed the above constructions of gates in small systems into arbitrary-many-qubit systems.
We could encode $n$ qubits using $2 (n+1)$ anyons with total charge 0, which form a dense topological quantum memory as in Fig.\ref{fig:groupfigure-01}(d). To read out the information of the $i$th qubit, we can map the $n$-qubit system to a single-qubit system  similar to that in Fig.~\ref{fig:groupfigure-01}(b), and then transmit the information on the $i$th qubit to an initialized qubit via {\it anyon-exchange gates} $E$ (See Fig.~\ref{fig:groupfigure-02})). The anyon-exchange gate can be carried out by braids (error rate $\sim 10^{-10}$ for braid length $\sim 100$ or less) as in~\cite{Xu08}, or by projective measurement (especially for k=2)~\cite{Xu092} similar to the measurement-only topological quantum computation~\cite{Bonderson08}, but here the measurement realizes {\it a gate instead of a single braiding operation}. 
In the other limit of non-dense encoding, we can instead use $4 n$ (or $3 n$) anyons to encode $n$ qubits separately, and then combine the information on, e.g., the first two of the qubits, into a two-qubit system composed of six anyons if necessary (see Fig.~\ref{fig:groupfigure-02}(b)).  More generically, using fixed-sized registers of dense memory provides a good approximation to maximum density while minimizing the area a braid encloses.  For example, using 8 anyons per 3 qubits is more efficient than the best non-dense encoding (3 anyons per qubit). Regardless, the gates in small systems can be readily embedded into arbitrary-many-qubit systems via anyon-exchange gates, no matter which encoding scheme one adopts.

\section{discussion}
In conclusion, we have developed a unified framework for topological quantum computation with anyons. We showed the equivalence between sparse and dense qubit
encoding schemes and exploited this equivalence to combine and decombine quantum information on different qubits via anyon-exchange gate, which enables us to embed gates in small systems into large systems. For universal topological quantum computation, we have given a unified construction of multi-qubit gates, especially the Toffoli gate. For all SU(2)$_k$ ($k\geq 2$) anyon models, we have reduced the difficulty of braid construction for multi-qubit gates to a bounded dimensionality (three or less), where anyon models that do not support universal computation may be made universal by adding non-braiding operations.

For more general anyon models we can use four non-Abelian anyons $a_{1,2,3,4}$ (which may have different topological charges), with fusion rules like $a_2 \otimes a_3=c_1 \oplus c_2 \oplus \cdots$, and with total topological charge $e$. The dimension of the four-anyon system might be larger than 2, in which case one can choose a two-dimensional subspace (expanded by basis states such as $|((a_1 (a_2 a_3)^{c_1})^d a_4)^e\rangle$ and $|((a_1 (a_2 a_3)^{c_2})^d a_4)^e\rangle$) to represent a single qubit. Then we can combine two single-qubit systems into a two-qubit system (such as $|( ((a_1 (a_2 a_3)^{c_{1,2}})^d (a_6 a_7)^{c_{1,2}})^f a_8 )^g\rangle$) using anyon-exchange gates.
If we choose the $a_{1-4}$ such that $c_{1}$ is Abelian while $c_2$ is non-Abelian, and can implement any qubit operation by braiding $a_{1}$ around two anyons with topological charges  $c_2$, the construction of multi-qubit gates in this paper is directly applicable. Otherwise, the dimension of searching space might not be reduced to that of a single qubit or less, and we need to revise the construction or use other methods.

\begin{acknowledgements}
The authors wish to thank L.~Hormozi and P.~Bonderson for helpful discussions.  This work was partially supported by the NSF through the JQI Physics Frontier Center.
\end{acknowledgements}

\end{document}